\documentclass[useAMS]{mn2e}
\title{Near-Infrared $H$ and $K$ band studies of the  2006 outburst of
the recurrent nova RS Ophiuchi}

\author[Banerjee et al]{D. P. K. Banerjee, R. K. Das and N. M, Ashok
\thanks{E-mail: orion@prl.res.in (DPKB); rkdas@prl.res.in (RKD); ashok@prl.res.in (NMA)}\\
Physical Research Laboratory, Navrangpura, Ahmedabad 380009, India\\}

\usepackage{graphicx}
\begin{document}

\pagerange{\pageref{firstpage}--\pageref{lastpage}} \pubyear{2008}

\maketitle

\label{firstpage}

\begin{abstract}

We present near-infrared photo-spectroscopy in the   $H$ and $K$ bands of the
2006 outburst of the recurrent nova RS Ophiuchi (RS Oph). The observations cover
the period between 1 to 94 days after the eruption. The near infrared light curve
 is presented. An extensive set of spectra are presented, lines identified and
 the  general characteristics of the spectra discussed. Analysis
of the HI line profiles show the presence of broad wings on both flanks of
a strong central component indicating  the presence of a bipolar velocity flow
in the ejecta. Such a flow is kinematically consistent with the  bipolar structure
that the object displays  in high-resolution spatial images. We discuss the
behaviour and origin of the Fe II lines at 1.6872 and 1.7414 $\mu$m  that are
prominently seen throughout the span of the observations. It is examined and
shown that  Lyman $\alpha$ and Lyman continuum fluorescence
are viable mechanisms to excite these lines. We draw upon the result, that collisional
excitation can also contribute in exciting and significantly enhancing the
strength of these Fe II lines, to propose that these lines originate from a
site of high particle density. Such a likely site could be the high-density,
low temperature contact surface that should exist in the shockfront  in between
the shocked ejecta and red giant wind. Recombination analysis of the HI lines
indicate deviations from Case B conditions during most of the span of our
observations indicating optical depth effects. It appears likely that the breakout
of the shockfront had not  yet occured till the end of our observations. An analysis is made of the  temporal evolution of the [Si VI] 1.9641 $\mu$m coronal line and another coronal line at  2.0894 $\mu$m which is attributed to [Mn XIV].  Assuming collisional effects to dominate in the hot coronal gas, estimates
are made of the ion temperature in the gas.
\end{abstract}


\section{Introduction}

The recurrent nova RS Oph underwent its most recent outburst on 2006 February
12.83. It was discovered at $V$=4.5 near the optical peak (Hirosawa 2006).
The earlier outbursts of RS Oph have been well documented - the previous
eruptions having occured in 1898, 1933, 1958 and 1985. The RS Oph binary
system comprises of a white dwarf primary accompanied by a red giant
secondary - the orbital period of the system being 455.72 $\pm$ 0.83 days (Fekel et al. 2000). A more recent study by Brandi et al. (2009) finds a closely similar
value of 453.6 $\pm$ 0.4 days for the orbital period.
There is some uncertainty regarding the exact classification of the secondary
companion - suggested classifications range from K0 III to M4 III with
M2 III being the most likely (Worters et al. 2008 and references therein). The
recent outburst has been studied in great details at several wavelength regimes
viz. in X-rays (e.g. Sokoloski et al. 2006, Bode et al. 2006), in the optical
(Iijima 2006, Skopal et al. 2008; Buil 2006, Fujii 2006), in the
infrared (e.g. Das et al. 2006, Evans et al. 2007a, Monnier et al. 2006) and
in the radio (e.g. O'Brien et al. 2006; Kantharia et al. 2007; Rupen et al. 2008).
In our previous study (Das et al. 2006), we had presented and analysed the
$J$ band near-infrared data. Our focus, at that stage, was on the IR detection and
behaviour of a shock wave that is generated as the high velocity ejecta from
the outburst interacts with  the pre-existing red giant  wind. The velocity
changes in the propagating shock front, notably its rapid deceleration, were
inferred by noting the large changes in the linewidths of the Pa $\beta$ 1.2818
${\rm{\mu}}$m and the Ly $\beta$ fluoresced OI 1.1287 ${\rm{\mu}}$m lines in the
$J$ band. In the present work, we study the general near-IR characteristics of
the present eruption in greater detail. To achieve this, we extend the analysis
of  some of the earlier $J$ band data but, more importantly,  present new and
extensive $H$ and $K$ spectroscopic and photometric observations of RS Oph upto
94 days after its outburst.

\section{Observations}

RS Oph was observed regularly since its discovery on 2006
Feb 12.83. Both spectroscopic and photometric observations in the near-IR $JHK$ bands
were obtained at the Mt. Abu 1.2m telescope. The spectra were obtained at  similar
dispersions of $\sim$ 9.75 {\AA}/pixel in each of the $J,H,K$ bands using the Near
Infrared Imager/Spectrometer with a 256$\times$256 HgCdTe NICMOS3 array. The log of
spectroscopic observations has been presented in Table 1. In each of the bands,
generally a set of at least two spectra were taken with the nova off-set to two
different positions along the slit (slit width $\sim$1 arcsec). The dithered spectra
were subtracted from each other to remove the contribution from the sky. In order to
remove telluric features in the object spectra, spectra of a standard star (SAO
122754; spectral type A0V)  were also recorded; the general attempt always
being to obtain the object and standard star spectrum at the same airmass.
The spectra were extracted  and analysed using the APEXTRACT task in IRAF and
wavelength calibration was done using a combination of
OH sky lines and telluric lines that register with the stellar spectra.  Following
the standard reduction procedure, the Hydrogen Paschen and Brackett
absorption lines were first removed by manual extrapolation from the spectra of the
comparison star. While the absorption lines are generally well removed in this process, it is estimated that residuals amounting to 5 percent of the line strength may still remain. The nova spectra were then divided by the spectrum of the
comparison star; these ratioed spectra were multiplied by  a blackbody curve
corresponding to the effective temperature of the comparison star to yield
the final spectra. The ratioing process, while removing telluric features
sufficiently well, leaves some residuals  in the wavelength regions where
telluric absorption is strong. This applies significantly to the  region around
2.0 ${\rm{\mu}}$m in the $K$ band which is  strongly affected by atmospheric
carbon-dioxide.

The $JHK$ photometry of RS Oph was done under photometric sky conditions using the
imaging mode of the NICMOS3 array. In each of the $J,H,K$ filters, several frames in
5 dithered positions, offset typically by $\sim$20 arcsec, were obtained of both the
nova and a selected standard star. Photometric reduction was done using the aperture
photometry task APPHOT in IRAF. The derived $JHK$ magnitudes, with typical errors in
the range of 0.01 to 0.03 magnitudes, along with the log of the photometric
observations are given in Table 2.

\begin{table}
\caption{A log of the spectroscopic observations of RS Ophiuchi. The date of outburst
is taken to be 2006 Feb 12.83 UT}
\begin{tabular}{lcccccc}
\hline\
Date & Days        & &         & Integration time &   \\
2006 & since        & &        & (sec)            &    \\
(UT) & Outburst  & & \emph{J} & \emph{H}         & \emph{K}    \\
\hline
\hline
Feb. 13.9929  & 1.1629   & & 10      & 10               & 10 \\
Feb. 14.9915  & 2.1615   & & 10      & 10               & 10 \\
Feb. 15.9898  & 3.1598   & & 10      & 10               & 10 \\
Feb. 16.9866  & 4.1566   & & 10      & 10               & 10 \\
Feb. 19.9759  & 7.1459   & & 10      & 8                & 10 \\
Feb. 20.9721  & 8.1421   & &  5      & 8                & 10 \\
Feb. 21.9463  & 9.1163   & & 10      & 10               & 15 \\
Feb. 22.9505  & 10.1205   & & 10      & 10               & 10 \\
Feb. 24.9705  & 12.1405   & & 10      & 10               & 10 \\
Feb. 26.9926  & 14.1626   & & 10      & 10               & 10 \\
Feb. 28.9542  & 16.1242   & & 15      & 10               & 10 \\

Mar. 02.9465  & 18.1165   & & 20      & 20               & 20 \\
Mar. 16.9465  & 32.1165   & & 20      & 30               & 30 \\
Mar. 17.8866  & 33.0566   & & 30     & 30               & 30 \\
Mar. 18.9372  & 34.1072   & & 30     & 30               & 30 \\
Mar. 19.8833  & 35.0533   & & 30     & 30               & 30 \\
Mar. 20.8733  & 36.0433   & & 45     & 30               & 30 \\
Mar. 22.8682  & 38.0382   & & 45     & 45               & 45 \\
Mar. 26.8778  & 42.0478   & & 60     & 60               & 60 \\
Mar. 27.8644  & 43.0344   & & 60     & 60               & 60 \\
Mar. 29.8462  & 45.0162   & & 75     & 90               & 90 \\
Mar. 30.8629  & 46.0329   & & 75     & 90               & 90 \\
Mar. 31.8701  & 47.0401   & & 90     & 90               & 90 \\

Apr. 02.8736  & 49.0436    & & 90     & 120               & 120 \\
Apr. 03.9691  & 50.1391    & & 90     &  120               & 120 \\
Apr. 05.9607  & 52.1307    & & 90     &  90                & 90 \\
Apr. 10.8936  & 57.0636    & & 90     & 120               & 90 \\
Apr. 12.8454  & 59.0154    & & 90     & 120               & 90 \\
Apr. 22.9392  & 69.1092    & & 90     & 90               & 90 \\
Apr. 29.8355  & 76.0055    & & 120     & 120               & 120 \\

May. 01.8149  & 77.9849    & & 120     & 180               & 180 \\
May. 03.810   & 79.980     & & 180     & 180               & 180 \\
May. 11.8353  & 88.0053    & & 270     & 270               & 270 \\
May. 17.8103  & 93.9803    & & 270     & 270               & 270 \\

\hline
\end{tabular}
\end{table}

\begin{table}
\centering
\caption{A log of the photometric observations of RS Oph. The date of outburst
is taken to be 2006 February 12.83 UT}

\begin{tabular}{lccccc}
\hline
Date & Days         & &             & Magnitudes&          \\
2006 & since        & &             &           &          \\
(UT) & Outburst     & & \emph{J}    & \emph{H}  & \emph{K} \\
\hline
\hline
Feb. 23.9518  & 11.1218  & & 4.82 & 4.54    & 3.85  \\
Feb. 27.9259  & 15.0959  & & 5.12 & 4.79    & 4.19  \\

Mar. 01.9485  & 17.1185  & & 5.30  & 4.97   & 4.42 \\
Mar. 21.8986  & 37.0686  & & 6.46  & 6.12   & 5.64 \\
Mar. 22.9645  & 38.1545  & & 6.43  & 6.14   & 5.63\\
Mar. 26.9534  & 42.1234  & & 6.67  & 6.23   & 5.75 \\
Mar. 27.9189  & 43.0889  & & 6.69  & 6.30   & 5.81 \\
Mar. 29.9696  & 45.1396  & & 6.73  & 6.32   & 5.81 \\

Apr. 29.9037  & 76.0737  & & 7.12  & 6.53   & 6.33 \\
May. 01.8913  & 78.0613  & & 7.16  & 6.56   & 6.43 \\
May. 03.8845  & 80.0545  & & 7.35  & 6.55   & 6.49 \\
May. 11.9199  & 88.0899  & & 7.57  & 6.92   & 6.56 \\
May. 17.9127  & 94.0827  & & 7.86  & 7.01   & 6.74 \\
\hline
\end{tabular}
\end{table}

\begin{figure}
\centering
\includegraphics[bb= 4 1 286 576, width=3.3 in,height=6.0 in,clip]{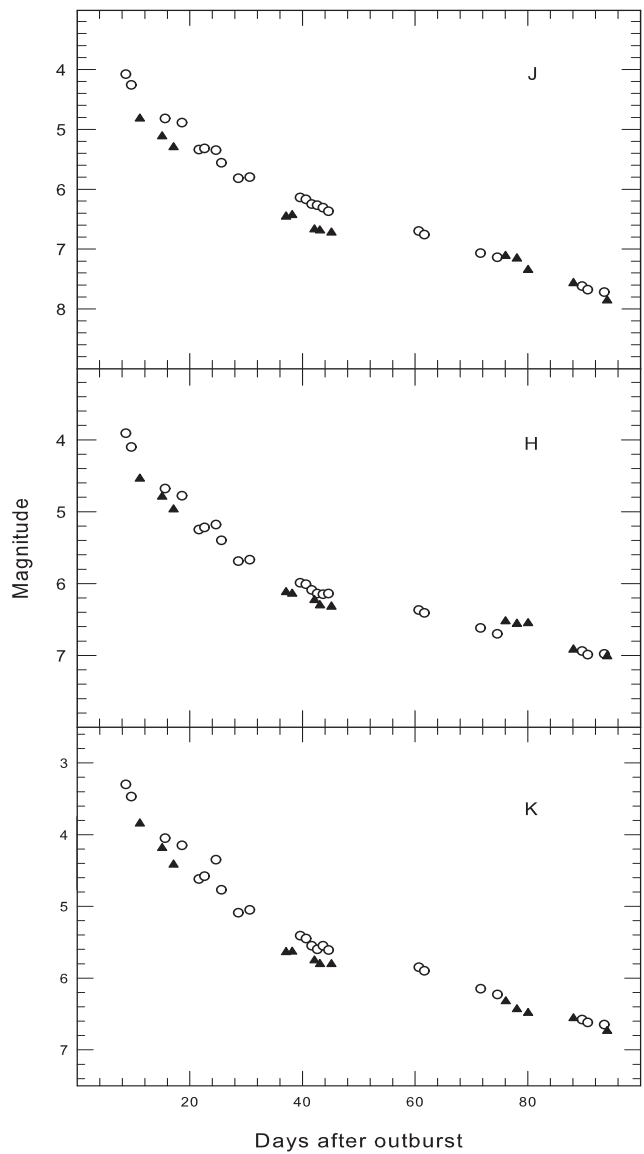}
\caption[]{ The $J, H, K$ lightcurves for RS Oph, based on data from Table 2,
is shown with filled triangles. To show the similarity of the  lightcurve
behaviour from different outbursts, the data from the 1985 outburst
 (Evans et al.  1988)  are also shown with  open circles.}
\label{fig1}
\end{figure}


\section{Results}
\subsection{Photometry: The $JHK$ Lightcurve}
We present in Figure 1 the $JHK$ lightcurves of RS Oph using photometric data
obtained between 11 to 94 days after outburst. Although we would have liked
to have sampled the lightcurve more frequently, our efforts were focused more
on the spectroscopy.
Also superimposed in Figure 1 is  the near-IR lightcurve of the 1985 outburst
which was obtained by Evans et al. (1988). While it is known that
the visual lightcurve behaves similarly from outburst to outburst (Rosino
$\&$ Iijima 1987), an equivalent comparison for the near-IR lightcurve
had not been possible earlier for lack of data.  The present data shows
that the IR decline is quite similar to that of the 1985 outburst.
However, some caution is needed while  comparing the near-IR light curves because of the effect of emission lines on the measured continuum flux in the photometric bands. The JHK spectra of RS Oph has prominent emission lines, some of which lie at the edge of the photometric bands. Filter transmissions in these regions can vary significantly in different photometric systems thereby contributing different amounts of the  line flux to the measured continuum. For example, Whitelock et al. (1984) point out the  significant effect of the HeI 1.083 $\mu$m line on the measured J magnitudes in emission line stars (the HeI line  lies at the edge of the J band and can be strong).

A compilation of several observations of the  pre-outburst or quiescent
photometry of RS Oph shows that it has a mean  pre-outburst values of $J$ = 7.72;
$H$ = 6.945; $K$ = 6.66 (Evans et al. 1988; Table 1 therein). From our data
we see that in  each of the $JHK$ bands the magnitude returned to its pre-outburst
value by $\sim$  88 days.  Our last observation on day 94 suggests  that
the decline appears to continue below this level - such a trend  was clearly seen
by Evans et al. (1988) for the 1985 outburst.

\subsection{Spectroscopy: General characteristics of the $JHK$ spectra and line
identification}
Although spectra of RS Oph were obtained on  several days, we present in
Figures 2  and 3 selected $H$ and $K$ band spectra that suffice to give the general
trend of the spectral evolution of the object. Since a
similar mosaic for the $J$ band spectra  has already been presented in Das et al.
(2006), it is not included here. However, to facilitate studying the overall
evolution of the
object in the $J$, $H$ and $K$ bands, we show three representative $J$ band spectra
in Figure 4 and also  expanded views of selected $H$ and
$K$ band spectra. These spectra in Figure 4 enable a  proper line identification
to be made  and to also bring out the weaker lines which are not easily discernible
in Figures 2 and 3. The line identification is given in Table 3 and
certain related aspects are discussed in more details below.

Over the span of our observations, the HI recombination lines remain prominent in the
$JHK$ spectra - these include Brackett $\gamma$ in the $K$ band, Br 10 to 21
in the $H$ band, and Paschen $\beta$ and Paschen $\gamma$ in the $J$ band.
Among the  HeI lines,  those at 1.0830, 1.7002 and 2.0581 $\mu$m are prominent.
The other weaker lines of HeI detected are those at  1.1969, 1.2527,
1.2968, 2.1120 and 2.1132 $\mu$m. The other prominent lines in the spectra
are the Ly $\beta$ fluoresced 1.1287 $\mu$m and continuum fluoresced
1.3164 $\mu$m OI lines respectively and two lines at 1.6872 and 1.7414 $\mu$m
which we attribute to Fe II. A more detailed discussion on these  Fe II lines
is made in a following subsection. Nitrogen  is  detected
as a broad feature centered around 1.245 $\mu$m which is seen conspicuously
in the spectrum soon after outburst (top panel of Figure 4). This feature,
which could be a blend of NI lines at 1.2461 and 1.2469 $\mu$m, weakens
with time but can still be readily discerned in the later spectra (second and
third panel from
top of Figure 4) on the blue wing of the HeI 1.2527 $\mu$m line. An
additional NI feature, a blend of the NI 1.2074 and 1.2096 $\mu$m lines, also
appears to be detected (however see Evans et al. 2007 who attribute this
feature to OII 1.2084 $\mu$m). Weak features of carbon  are also possibly being
detected. Immediately after outburst (top panel of Figure 4), a broad conspicuous
bump is seen centered around 1.175 $\mu$m. This feature could be
a blend of several carbon lines that lie between 1.1653 and 1.1886 $\mu$m
(the strongest of these are the  1.1748, 1.1753 and 1.1755 $\mu$m lines - the
blended feature of
these lines appears to persist weakly at 1.1748 $\mu$m for quite some
time after outburst ). These carbon lines which are weak here,  can be
extremely strong in novae which have  less massive central white dwarfs
(CO novae) e.g. in V1280 Sco (Das et al. 2008) and V2274 Cyg (Rudy et al. 2003).
In this context it is noted that the analysis of optical and X-ray spectra
indicate the ejecta to be enriched in nitrogen and depleted in carbon (Ness et
al. 2008 and references therein).

Among the coronal lines,  we detect [Si VI] 1.9641 $\mu$m, [S IX] 1.2520   $\mu$m,
[Al IX] 2.0444 $\mu$m and a line at $\sim$ 2.09 $\mu$m which could possibly be [Mn XIV]
2.0887 $\mu$m. The [Si VI] 1.9641 $\mu$m line appears at the edge of our
instrument spectral window i.e. in a region of low instrument and atmospheric
transmission. The spectrum in this region is noisy making it difficult to accurately estimate  the strength of the line. However we can state, with a fair level of
confidence, that the line begins to be clearly detected in the spectrum of 26.88 March 2006.
This sets an upper limit on when the line began to be seen
in the $K$ band.  On the other hand, it is difficult to be definitive
about when the [S IX] 1.2520 $\mu$m line first begins to appear because
it is severely blended
with the HeI 1.2527 $\mu$m line. But from an examination of all our $J$ band
spectra (Das et al. 2006), we see a significant strengthening of the
1.2527 HeI plus [S IX] feature between 20-29  March. It is likely that
this strengthening is due to the emergence of the [S IX] 1.2520 $\mu$m line
at around this time. Though the identification is uncertain, we attribute the
observed line at  $\sim$ 2.09 $\mu$m to  [Mn XIV] 2.0894 $\mu$m. The [Mn XIV]
2.0894 $\mu$m line is
not one among the commonly observed coronal lines in novae. Wagner $\&$ Depoy
(1996) did observe a  line at 2.092 $\mu$m in nova V1974 Cyg during its coronal
phase and tentatively assigned it to [Mn XIV] 2.0894 $\mu$m. An analysis of the temporal evolution of the coronal lines is given in section 3.7.

There is no evidence for dust formation in the ejecta till the end of
our observations. Neither the $JHK$ magnitudes
show a development of an infrared excess nor do the slope of the continua in
the spectra show an upturn towards longer wavelengths indicative of dust
formation. It may be noted that Spitzer observations obtained over the period
 208-430 days after the eruption revealed evidence for silicate dust features
at 9.7 and 18 $\mu$m (Evans et al. 2007b). But this silicate dust appears not to
have formed in the nova ejecta; rather it has been  interpreted as being
a part of the redgiant wind that existed prior to the 2006
eruption and has survived the outburst (Evans et al. 2007b).

All the emission lines in the $H$ and $K$ bands (Figures 2 and 3) show  a
rapid narrowing with time similar to what was earlier reported for  the
Pa $\beta$ 1.2818 $\mu$m and OI 1.1287 $\mu$m lines in the $J$ band (Das et al.
2006). Such behaviour was interpreted in terms of the deceleration of the
shock wave as it propagated through the pre-existing red giant wind. Since
a detailed analysis of the evolution of the shock, from line profile analysis,
has  already been presented for the $J$ band lines, we do not repeat a similar
analysis for the emission lines seen in the $H$ and $K$ bands. However,  we
may briefly mention that the Br $\gamma$ line was similarly analysed
as the $J$ band lines and its line velocity shows a similar temporal
evolution as the Pa $\beta$ line.

\begin{table}
\caption[]{List of observed lines in the $JHK$ spectra}
\begin{tabular}{llrr}
\hline\\
Wavelength         	& Species             	& Other contributing& \\
(${\rm{\mu}}$m)    	&                     	& lines $\&$ remarks& \\
\hline
\hline \\

1.0830            & He\,{\sc i}      & 				&\\	
1.0938   	   	& Pa $\gamma$     	&        			& \\
1.1126   		& Fe II             &  				&\\
1.1287   		& O\,{\sc i}    	& 				&\\
1.1626  		& He\,{\sc ii}    	& 		&\\
1.1748   		& C\,{\sc i}        & Possible blend of several CI&\\
         		&                   & lines between 1.1653 $\&$ 1.1886&\\
1.1969   		& He\,{\sc i}      &  				&\\
1.2074            & N\,{\sc i}        &blended with N\,{\sc i} 1.2096     & \\
1.2470  		& N\,{\sc i}       &  blended with N\,{\sc i} 1.2461       &\\
1.2527  	      & He\, {\sc i}     &           &\\
1.2520            & [S\, {\sc ix}]     &            & \\
1.2818   		& Pa $\beta$       	& 				&\\
1.2968  		& He\,{\sc i}?                 &  				&\\
1.3164   		& O\,{\sc i}		&                               &\\

1.5133            & Br21                 &               &\\
1.5184   		& Br20                 &               &\\
1.5256   		& Br19                 &   			&\\
1.5341   		& Br18              	&  				&\\
1.5439   	   	& Br17              	&  				&\\
1.5557   		& Br16              	&   				&\\
1.5685  		& Br15             &    				&\\
1.5753   		& u.i.              &  unidentified				&\\
1.5881   		& Br14             & 				&\\
1.6109    		& Br13    		    &  				& \\
1.6407  		& Br12       		&               & \\
1.6806   		& Br11    	    	&  				& \\
1.6872   		& Fe\,{\sc ii}   	&  				& \\
1.7002   		& He\,{\sc i}   	&  				& \\
1.7362  		& Br10      		&       		& \\
1.7413   		& Fe\,{\sc ii}      &  				&\\

1.9446   		& Br8      	& 				&\\
1.9641   		& [Si\,{\sc vi}]      	& 				&\\
2.0444   		& [Al\,{\sc ix}]  	& 				&\\
2.0581   		& He\,{\sc i}  	&				&\\
2.0894   		& [Mn\,{\sc xiv}]	& uncertain identification				&\\
2.1120   		& He\,{\sc i} 	&blended with He\,{\sc i} 2.1132	&\\
2.1369   		& u.i 			&	unidentified			&\\
2.1655   		& Br $\gamma$   &  				&\\

\hline
\end{tabular}
\end{table}

\begin{figure*}
\centering
\includegraphics[bb=2 1 689 766,width=7.0in,height=5.0in]{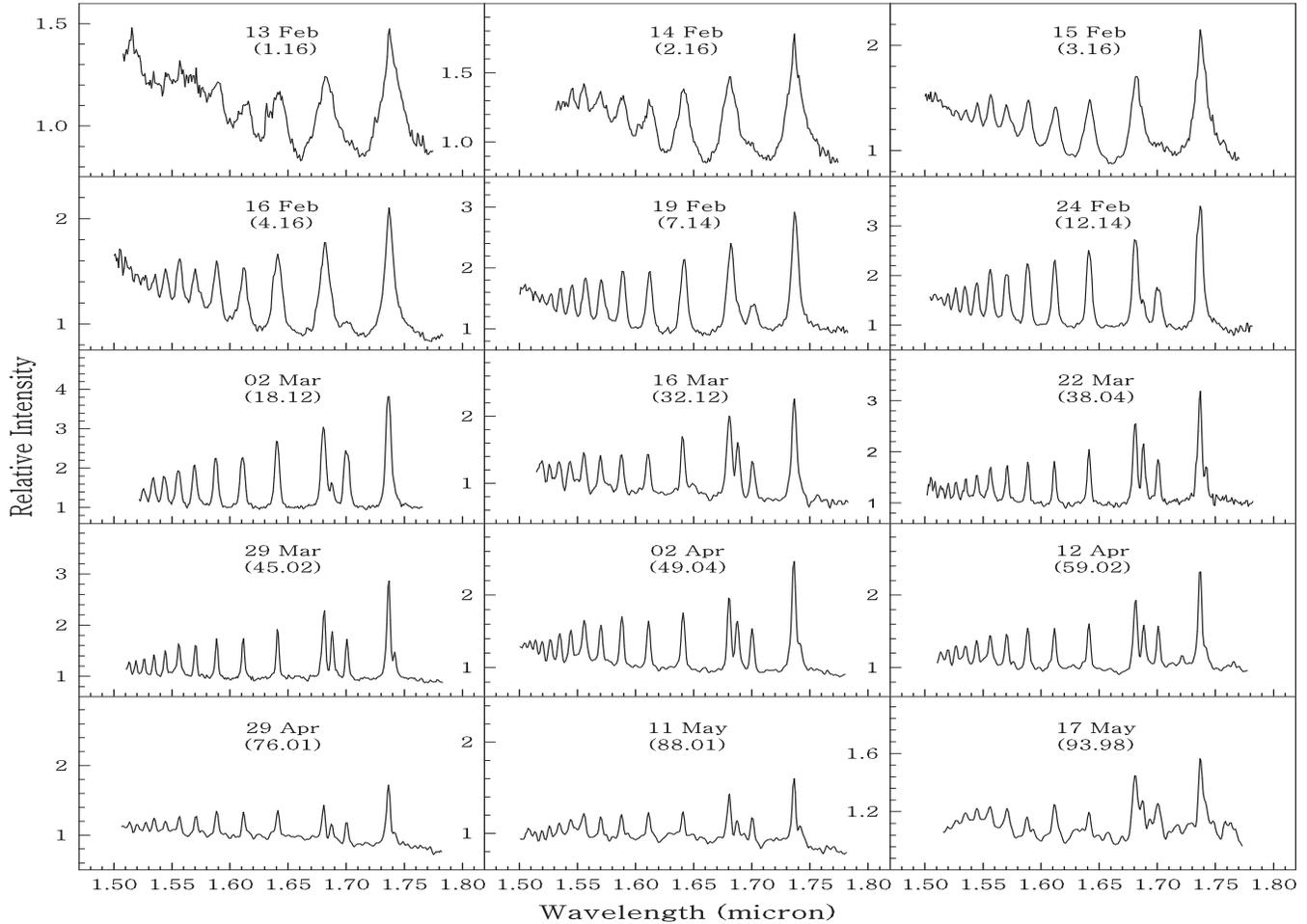}
\caption[]{ The $H$ band spectra of RS Ophiuchi on different days with the
flux normalized to unity at 1.65 ${\rm{\mu}}$m. The number inside the brackets represents the days elapsed since outburst.}
\label{fig2}
\end{figure*}

\begin{figure*}
\includegraphics[bb=1 1 691 763,width=7.0in,height=5.0in]{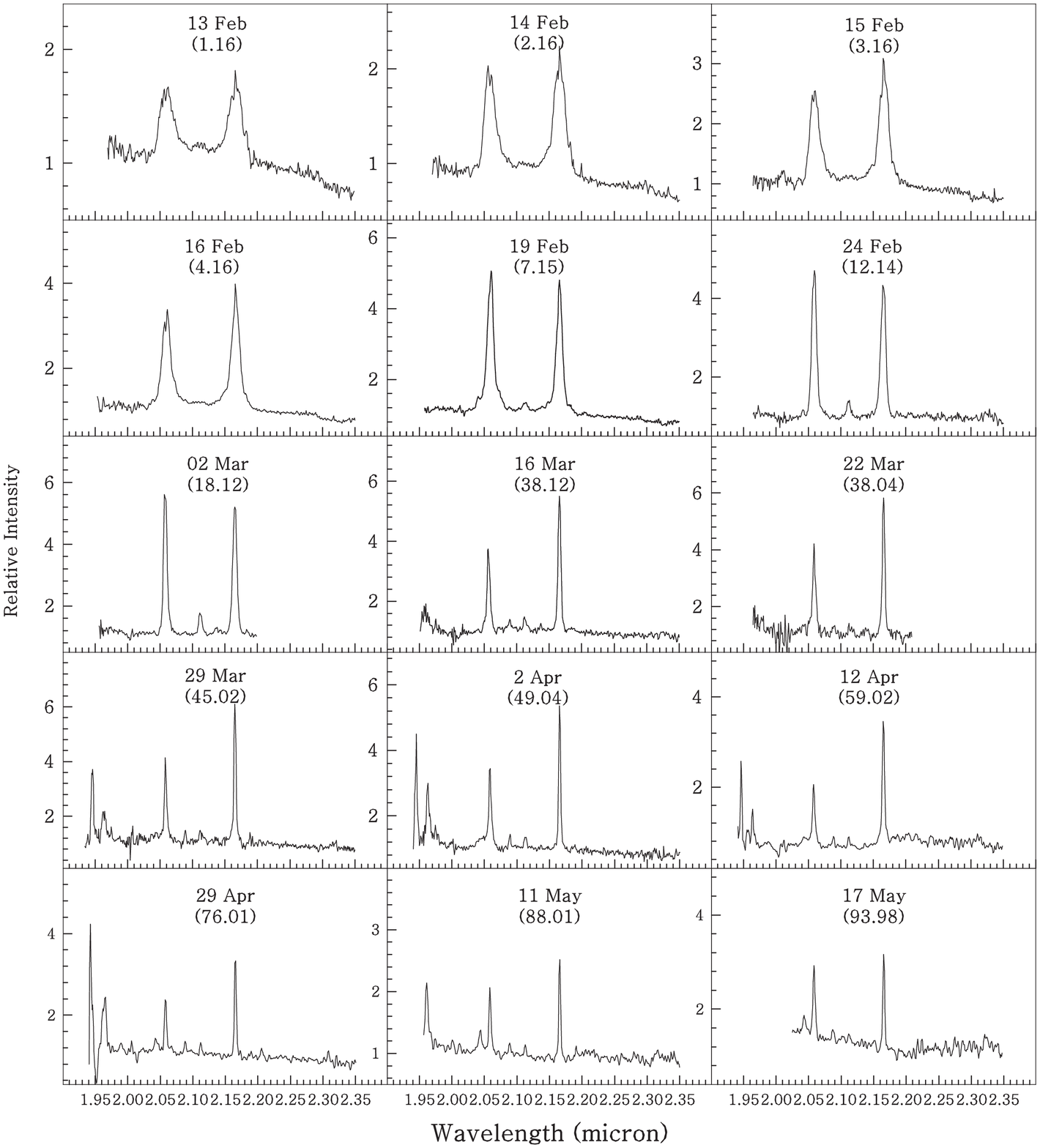}
\caption[]{ The $K$ band spectra of RS Oph on different days with the flux
normalized to unity at 2.2 ${\rm{\mu}}$m. The presence of  noise and weak
features at $\sim$ 2.0 ${\rm{\mu}}$m in some of the spectra are residuals
from incomplete  telluric subtraction as discussed in Section 2. The number inside
the brackets represents the days elapsed since outburst.}
\label{fig2}
\end{figure*}

\begin{figure}
\centering
\includegraphics[bb=2 1 282 976,width=3.5in,height=7.0in,clip]{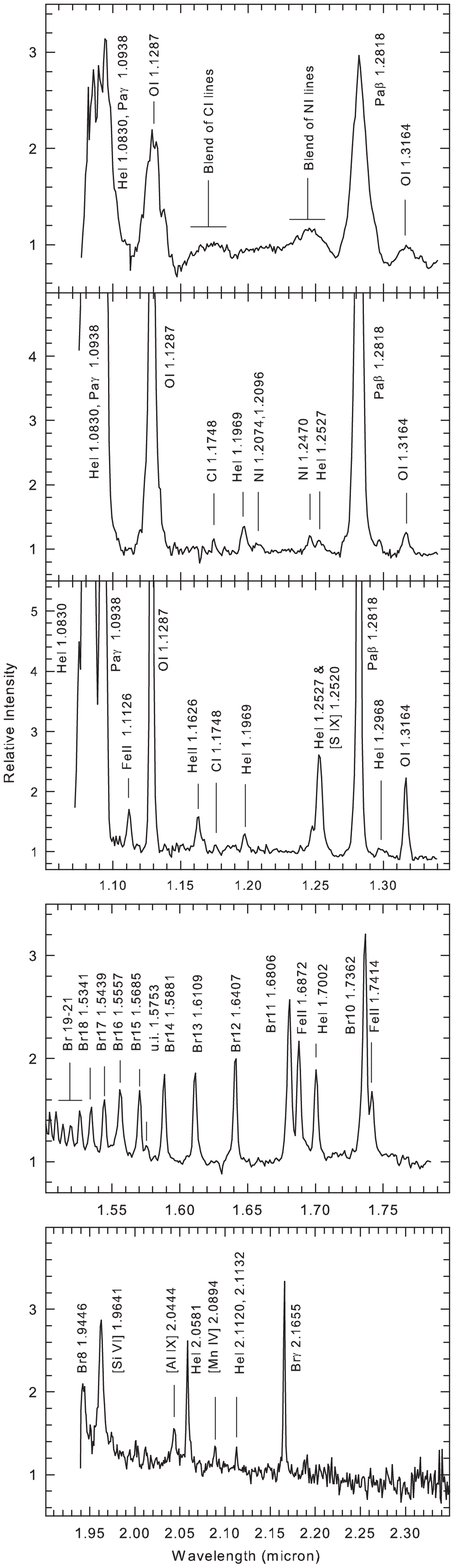}
\caption[]{ Identification of the observed lines in the $J,H$ $\&$ $K$ bands.
The top three panels show $J$ band spectra for 14 February, 24 February and
29 March 2006 respectively. The bottom two panels show the $H$ and $K$ band
spectra of  29 March and 11 May 2006 respectively ( the $K$ band spectra is
at bottom). Further details are provided in section 3.2 and Table 3.
}
\label{fig4}
\end{figure}

\subsection{Evidence for a bipolar flow}
A significant  finding  that has emerged from studies of the present outburst
of RS Oph is the presence of extended structure in the ejected material.
Images from the Hubble Space Telescope, taken 155 and 449 days after outburst
in the [OIII] 5007$\AA$   and [NeV] 3426$\AA$ lines, clearly show an
expanding hourglass shaped nebula around RS Oph (Bode et al. 2007, Harman
et al. 2008). Such bipolar structure is commonly encountered in planetary
nebulae and explained on the basis of the  ejecta interacting with a
non-uniform  circumstellar environment. Alternatively, the ejection of the
material could be intrinsically anisotropic but this scenario does not
appear to be favored here (Harman et al. 2008). In the first scenario, if
the pre-existing circumstellar
material has a density enhancement in the equatorial plane then the
outflowing ejecta is impeded from expanding in the equatorial region
while  expanding relatively more freely in the polar direction.This leads
to a constriction of the nebula in the equatorial region thereby giving
rise to an hourglass shape. Kinematically, this would imply that
the matter in the poles would flow out with a high velocity relative
to the matter in the waist of the hourglass. Evidence corroborating such a
bipolar velocity flow was reported in interferometric studies by Chesneau
et al. (2007) and from H$\alpha$ line profile analysis by Skopal et al. (2008).
By modeling the $K$ band visibilities obtained 5.5 days after outburst,
Chesneau et al. (2007) find evidence for two radial velocity fields - a slow
expanding ring like structure (likely the waist of the nebula)
with a velocity $\le$ 1800 km/s and a faster structure (possibly originating
from the polar regions) with a velocity of $\sim$ 2500-3000km/s. Skopal et al.
(2008) obtain a similar result from their H$\alpha$ profiles,
taken early after the outburst, which shows a main  emission component
flanked by two weaker high velocity components in the wings (Figure 2 of
Skopal et al. 2008).

Similar evidence for a bipolar flow is also seen from our near-IR data.
To illustrate this, we present in Figure 5,  representative profiles of the
Pa $\beta$ and Br $\gamma$ lines on Feb 16.987. As can be seen, the profiles
have a  strong central component flanked by two weaker components. In both
panels of Figure 5, we have fitted the profile with three
gaussians - a central gaussian for the core emission component and two
gaussians for the satellite components in the wings (the gaussians are shown by
dotted lines, their sum by the gray line and the observed data
by the continuous black line). It is seen that a three component gaussian
fits the data reasonably well. The fits indicate the presence of two high-velocity
components at radial velocities of -2540 and +2485 km/s for Pa $\beta$
and at -2340 and 2160 km/s for the Br $\gamma$ lines respectively (these
high velocity components appear to be associated with the faster structure
reported by Chesneau et al. 2007). The central components have FWHM's of
$\sim$ 2200 and 1770 km/s for the Pa $\beta$  and Br $\gamma$ lines
respectively.  We can interpret the results of Figure 5 as follows viz.
the core emission
can be associated with the slower expanding material from the waist of the
bipolar ejecta while the higher velocity satellite components are associated
with the flow from the polar regions.  It may be mentioned that Skopal et al. (2008)
draw a similar interpretation from their H$\alpha$ profiles.

\begin{figure}
\centering
\includegraphics[bb=4 25 295 456,width=3.4in,height=5.5in,clip]{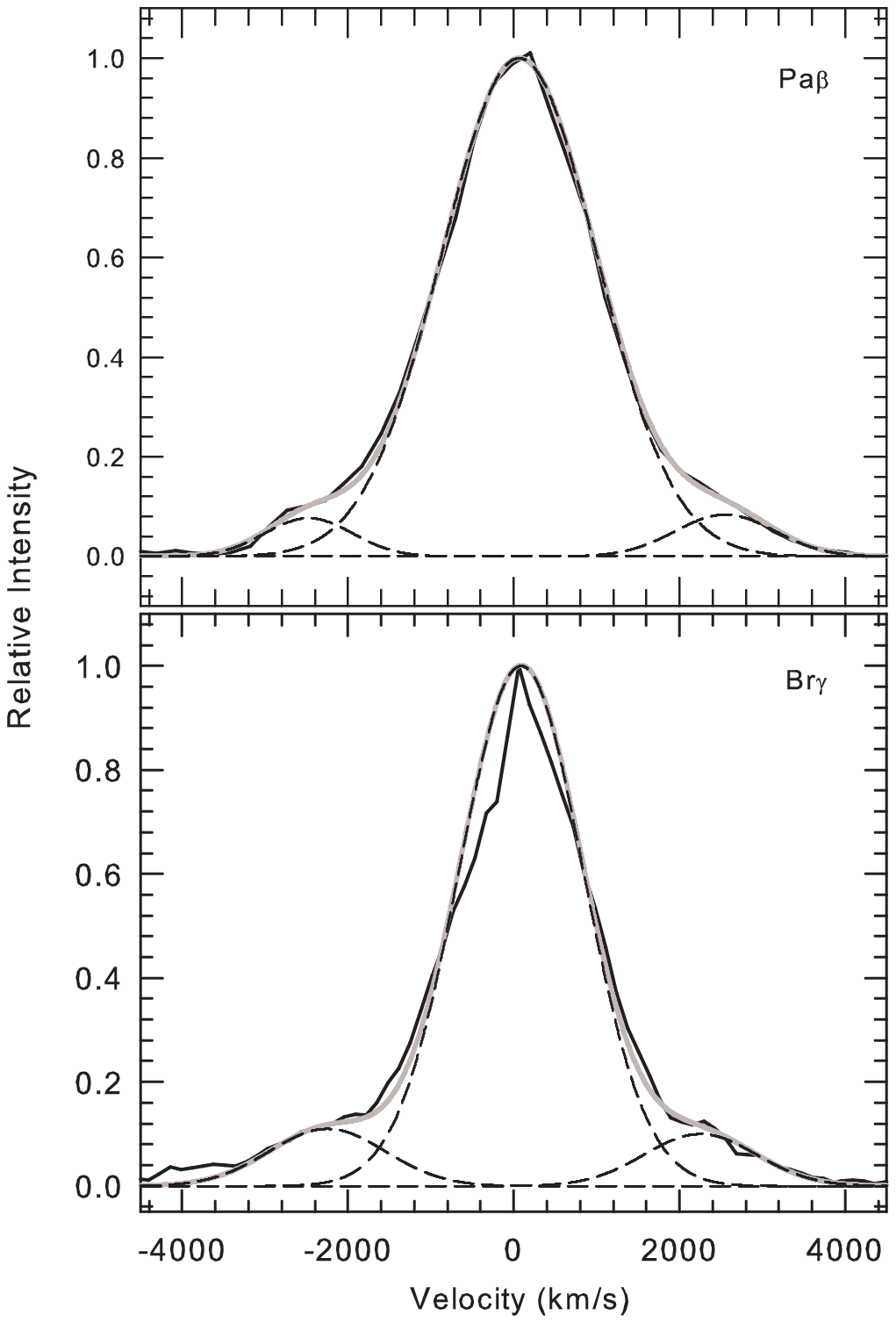}
\caption[]{ Line profiles of the
Pa $\beta$ and Br $\gamma$ lines on Feb 16.987 showing broad wings indicative of
a bipolar flow. A multigaussian fit of the profiles is shown - a gaussian for the
central component and two gaussians for the satellite components in the wings. The
gaussians are shown by dashed lines, their co-added sum by the gray line and the
observed data by the continuous black line. Further details are provided in
Section 3.3.}
\label{fig2}
\end{figure}

\subsection{The  Fe II lines at 1.6872 and 1.7414 $\mu$m}

One of the more uncommon spectral features that are seen in our $H$ band spectra
are prominent  lines of Fe II at 1.6872 and 1.7414 $\mu$m. In addition,  also present
in  our $J$ band spectrum of RS Oph is  the 1.1126 $\mu$m Fe II line. This
line constitutes one of the  so-called "one micron Fe II lines" that are
detected at  0.9997, 1.0171, 1.0490, 1.0501, 1.0863, and 1.1126 $\mu$m  in novae
(Rudy et al. 1991, 2000).  The 1 micron
lines,  termed so by Rudy et al. (1991) because of their wavelength location,
were first identified by Johansson (1977) in the spectrum of $\eta$ Carinae
and later seen  in a variety of  emission-line stars (Johansson $\&$
Jordan 1984; Hamann $\&$ Persson 1989; Rudy et al. 1991; Hamann et al. 1994)
and in AGNs (Rudy et al. 2000, 2001; Rodriguez-Ardila et al. 2002). In
comparison, the $H$ band Fe II lines are rather uncommon, at least in novae
spectra. In the case of novae, two of the earlier recorded instances of these
lines  are in the slow nova V2540
Ophiuchi  (Rudy et al 2002) and possibly also in the recurrent nova  CI Aquila
(Lynch et al. 2004). One of the main processes for the excitation of the Fe II
lines is considered to be Lyman $\alpha$  (Ly $\alpha$) fluorescence. Additional excitation mechanisms are Ly continuum fluorescence and collisional excitation. Collisional excitation, which can play a significant role in enhancing the strength of the FeII lines (Bautista et al 2004), is discussed in greater details in section 3.5. We first consider  the Ly alpha fluorescence process for the 1 micron lines. These lines specifically come from a three step process viz.  the line proper comes
from the decay of the $3d^5$$4s^2$$b^4G$ term at
$\sim$ 6.72 eV above the ground state; this term being fed as a second step
from decay of a $^4G^o$ term at $\sim$ 13.5eV  which is originally pumped in the first
step by Ly $\alpha$ fluorescence (Johansson $\&$ Jordan 1984).\\

In this context, we examine  the viability of a similar Ly $\alpha$ excitation mechanism for the
1.6872 and 1.7414 $\mu$m lines (it appears that the excitation mechanism for
these lines has been studied relatively less compared to the 1 micron
lines).  Using the  atomic line database of Kurucz (http://cfa-www.harvard.edu/amp/ampdata/kurucz23/sekur.html) we identify, at the relevant wavelengths,   two Fe II lines  at  1.6872 and 1.7414 $\mu$m
which can undergo a similar three step Ly $\alpha$ excitation process. The
1.7414 $\mu$m line results from the decay of the  $3d^6(^3_2F)4s$ $c^4\!F$ term at $\sim$ 6.222 ev above
the ground state; the 1.6872 $\mu$m line  from the
$d^5$$4s^2$\hskip 1mm$^2G$ term at 9.743 eV.  It so happens that both these
upper levels of the 1.6872 and 1.7414 $\mu$m lines can be fed by not just
one, but in fact by several high lying levels - each of these high lying
levels capable of being  pumped by Ly $\alpha$ photons. As examples for the
1.6872 $\mu$m line we note that photons at 1215.691 and 1215.707$\AA$, which
are closely coincident with the Ly $\alpha$ line center at 1215.671$\AA$, can excite
transitions from low lying levels in Fe II (at around 3.2 to 3.4 eV above ground
state) to the  higher excited levels  (at around 13.4 to 13.6 eV ). These  higher
levels can then decay via ultraviolet photons (at 1722.607 and 1683.961
$\AA$  respectively)  to the upper level of the 1.6872 $\mu$m transition (viz.
the $d^5$$4s^2$\hskip 1mm$^2G$ term) . Similarly for the 1.7414 $\mu$ line,
photons at 1215.691 $\AA$ and 1215.873 $\AA$  can Ly $\alpha$  fluoresce from
low lying levels in Fe II  to the  higher excited levels.
These  higher levels can then decay via UV photons (at 3370.341      and
3435.801 $\AA$ respectively) to the upper level of the 1.7414 $\mu$m transition
(viz. the  $3d^6(^3_2F)4s$ $c^4\!F$ term). Since HI  lines in novae (Ly $\alpha$
included) are routinely broad with  widths extending upto a few thousands of
km/s,  additional Fe II transitions at wavelengths away from the Ly $\alpha$
line center could
also  contribute to the Ly $\alpha$  fluorescence process. It is noted from the
Kurucz data, that there are indeed quite a few such lines - within a few Angstroms
of the Ly $\alpha$ line center (1 Angstrom corresponds to about 250 km/s at the
Ly $\alpha$ wavelength) - that could also contribute to the Ly $\alpha$
fluorescence process. It thus appears that Ly $\alpha$ fluorescence is a viable
mechanism  for the excitation of the  1.6872 and 1.7414 $\mu$m lines.

In addition, it also needs to be examined whether  Lyman continuum (Lyc)
photons can also play a role in exciting these lines. If the nova ejecta
is optically thick in Ly $\alpha$, as is expected, then it should be optically
thick in Lyc photons too - hence Lyc  fluorescence could also take
place. We find that this is indeed   viable  since several
transitions of Fe II can be identified   which are
excited by Lyc photons to high-lying energy levels and subsequently followed
by decay, via emission of an UV photon, to the upper levels of the 1.6872
and 1.7414 $\mu$m lines (one of several examples is the Fe II transition at
914.846$\AA$ between lower and upper states at 0.083 eV and 13.636 eV respectively
which can be 
the subsequent decay  of the higher excited state via a 1672.578$\AA$ line
populates the upper level of the  1.7414 $\mu$m line).

Rudy  et al. (2002) assigned the observed $H$ band lines with FeII lines at
1.6873 and 1.7414 $\mu$m. We would agree with the identification of the
1.7414 $\mu$m line  since a mechanism exists to account for its presence
viz. Ly $\alpha$ and Lyc fluorescence as has been discussed. However, there
is a possibility of an error in the identification of the 1.6873 $\mu$m line.
It is seen that actually there are two closeby lines of Fe II in this
region viz at 1.68732 and 1.68719 $\mu$m respectively (air wavelengths).
Identification of the observed feature with the 1.68719 $\mu$m (we refer to
this as the 1.6872 $\mu$m line throughout the text) appears correct since this
line, as demonstrated, can be excited by Ly $\alpha$  fluorescence. On the other
hand the 1.6873 $\mu$m line, proposed by Rudy et al. (2002) arises in emission
as a transition between the $(^3F)$ $4s$ $c^4F$ and  $(^5D)$ $4p$ $z^4F$ states
at $\sim$ 6.219 and 5.484 eV respectively. We are unable to identify any transition,
arising from Ly $\alpha$ fluorescence, that  can feed the upper level of this line
directly or in a secondary step (as in the case of the other observed Fe II
lines discussed so far). It therefore appears that the correct identification
of the observed feature should be with the 1.68719 $\mu$m line.

\subsection{Site of the near-IR line emission and studying the possibility of
shock breakout}
We study whether it is possible to identify the region in the ejecta
from where the Fe II and the other near-IR emission lines arise.
The excitation mechanism of the Fe II lines could offer a clue
in this matter. While Ly $\alpha$ fluorescence could be a vital or possibly
even the central process in exciting the Fe II lines, it may not be the sole
mechanism  and collisional excitation also  plays a significant role.
Such an inference is suggested from observations which show the absence, or
greatly reduced  strength from expected values, of the cascade lines that feed
the upper levels of the 1 micron lines (Bautista et al. 2004 for a detailed
discussion of the observational evidence). Bautista et al. (2004)
show that Fe II has several high-energy pseudometastable levels and that
collisional coupling between such metastable levels and radiative-emitting
levels is considerable. They show that electron-impact-induced transitions
from metastable levels strengthens lines throughout the  emission spectrum and in
particular  strengthens the lines that result from secondary decay
after Ly $\alpha$ pumping like the 1 $\mu$m lines. Model calculations of the
emissivities (Bautista et al. 2004; Figure 1 therein) show that the peak emissivities
of lines can be enhanced by a factor of ten or more when collisional transitions are
taken into account vis-a-vis when these transitions are  neglected in the
calculations. The emissivities of the lines are found to be dependent strongly
on the electron density and peak emissivities are seen to occur at high
densities in the range $10^{10}$ to $10^{12}$   electrons/cm$^3$.
The need for  high particle densities  to make these lines prominent is not
surprising - it is under such conditions that the number of collisional excitations
from the metastable levels can be expected to be enhanced. In essence, the model
calculations suggest that if the  1.6872, 1.7414 $\mu$m lines or the 1 $\mu$m lines
are strong, then high density conditions are likely to be prevalent in the ejecta.
This in turn, could offer an insight into the site of the Fe II emission. From an
inspection of Figures 2 and 3 it is seen that the 1.6872 and 1.7414  $\mu$m
lines remain prominent throughout the entire span of the observations. Since both
lines flank the redward wings of Br 11 and Br 10 lines respectively, they are
difficult to resolve as separate lines - even if they are present - during the
early phases of the expansion when all the emission lines are broad. However,
in spite of this, the presence of the 1.6872 $\mu$m line can be seen as a discernible
inflection on the red wing of Br 11 in the 19 February spectrum.  We interpret this
to mean that high density conditions prevail from almost the beginning to the end
of the observations
in  mid-May 2006 inspite of the considerable reduction in density that should be
expected as the ejecta expands (in case of a geometric dilution with time i.e.
$n_e$ proportional $r^{-2}$, the electron density is expected to  drop by a factor
of around 250 to 1000 between Feb 15-19 to our last observation on 17 May).\\

A possible site, where such high density conditions can be generated and sustained as long as the shock lasts,  is the region of discontinuity that is formed in a shock when a fast wind flows into a slow wind. Recent one dimensional hydrodynamical models have been  developed by Vaytet et al. (2007) to study the structure and evolution of such a  shock  in the RS Oph system. Their models aim to revise and improve  earlier studies (O'Brien, Bode and Kahn, 1992) by taking into account the ejection of material in the outburst as well as allowing the duration of the fast wind phase to be varied, as opposed to the instantaneous release of pure energy employed in their previous Primakoff-type models.  In addition, Vaytet et al (2007) also include the effects of  radiative cooling from the shocked gas which can significantly affect the dynamics of the system. Their model calculations show a complex structure for the shock front revealed through the presence of both forward and reverse shocks, with a separating  contact discontinuity. In both their updated models viz. the adiabatic wind model and the radiatively cooled model, the contact discontinuity is found to be the site of high density and low temperature conditions (relative to the material in the forward and reverse shocked regions). These effects (enhancement of density and lowering of temperature) are more pronounced in the radiatively cooled model.\\

In view of the above, we thus propose that the Fe II
emission, which is apparently favored by high-density conditions, originates from
a region associated with the contact discontinuity. In addition, Fe II is an ion of low
ionization stage implying  that it should originate in  a  zone of low kinetic
temperature (a similar reasoning applies for the presence of neutral OI
lines and suggests it coexists with Fe II). If our premise of associating the  Fe II
emission with the dense matter at the contact discontinuity  is correct, then the fact
that prominent Fe II emission is seen even upto $\sim$ 90 days after outburst  would
imply that the shockfront  had not broken out even at that stage. When is breakout
expected? The answer to this is complicated because of the bipolar velocity flow
in the nebula. Since the polar flow is much faster than the equatorial flow
(Figure 5, section 3.3), it should breakout much earlier. On the other hand, again
from Figure 5, there appears to be much lesser contribution to the line luminosity
from the high-speed bipolar jets vis-a-vis that from the equatorial flow. Thus even
if a breakout takes place in the polar direction and contribution to Fe II emission
from this region ceases, the bulk of the Fe II emission will still be expected from
the shockfront in the equatorial plane that has still to  break out. In general,
breakout will occur when the shockfront traverses a distance equal to the  finite
extent of the redgiant wind - this equals the product of the
velocity of the red giant wind and the time elapsed since the 1985 outburst (i.e.
21 years).   The velocity of the red giant wind has been estimated to be in the range
10-20 km/s (Gorbatski 1972). For this velocity range and  adopting a mean value of
1000 km/s for the shockfront velocity in the equatorial plane (from the FWHM values
reported in Das et al. 2006), breakout in the equatorial direction  could be expected
between 75-150 days. This timescale is consistent with our conclusion that breakout
may not have occured till
even $\sim$ 90 days after outburst. The break-out time estimates are susceptible to
errors because of uncertainties in the parameters involved (viz. the velocities of
the red giant wind and the shockfront). However, recombination analysis of the HI
lines in the following section, are reasonably in line with the conclusions drawn here.

\begin{figure*}
\centering
\includegraphics[bb=3 195 568 774,width=5 in,height=5.2in,clip]{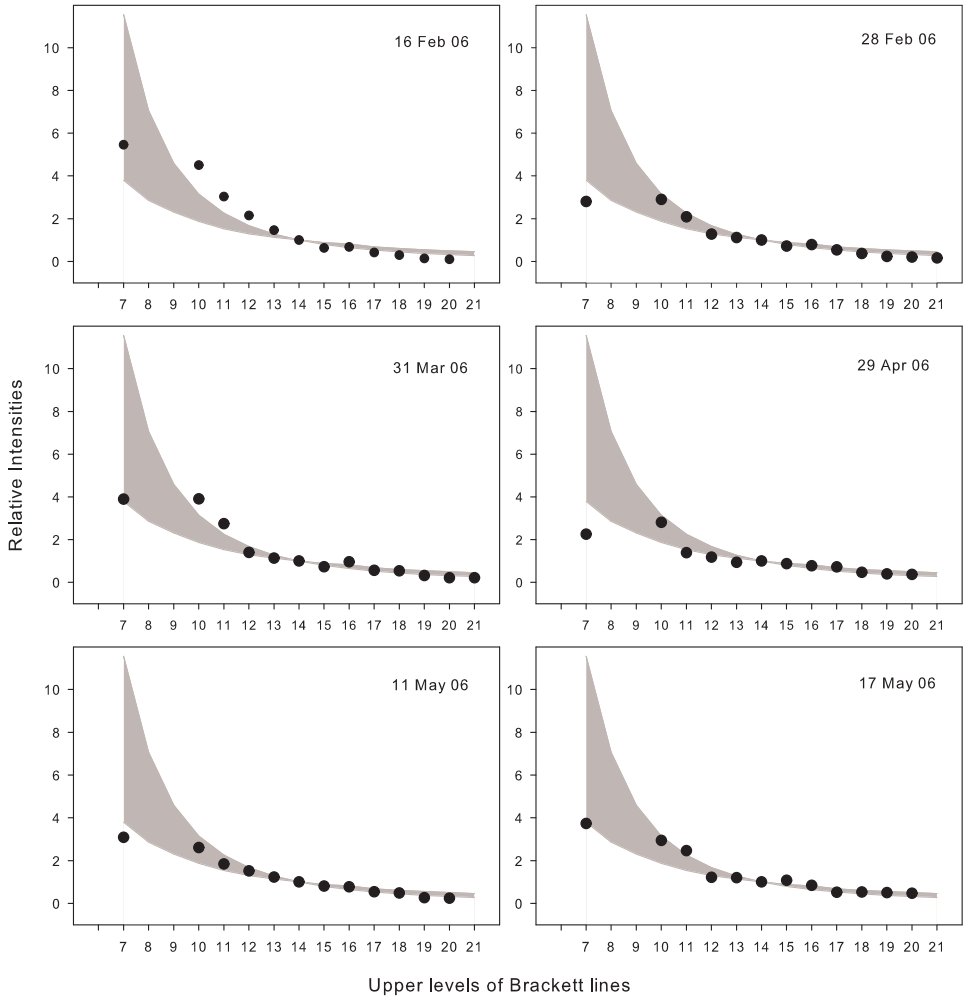}
\caption[]{ A comparison of the observed Brackett line intensities on different days
with Case B predictions. The shaded area in each panel indicates the range over which the Case B
model values are spread for the electron density $n_e$ varying between $10^5$
to $10^{12}$cm$^{-3}$ and temperature $T$ between 5000K to 20000K. The Br lines have been
normalized to unity at Br 14, for both Case B values and also for the observed data.
The error in the strengths of the  Br 10 and 11 lines is $\sim$ 30
percent and  in the range of 10 to 15 percent for the other lines.
Further details are given in section 3.6.}
\label{fig2}
\end{figure*}

\subsection{Recombination analysis of the HI lines}.

We present in Figure 6 the recombination Case B analysis for the HI lines on
six reasonably equispaced epochs that sample the evolution upto 94 days after the
outburst. The observed spectra were flux calibrated using the $JHK$ magnitudes of
Table 2 after correcting for reddening using $E(B-V)$ = 0.73 (or $A_v$ = 2.26;
Snijders 1987). The observed line luminosities were then compared with
the Case B emissivities of Storey $\&$ Hummer (1995). Three points may be
made regarding Figure 6. Since Case B calculations are only
mildly sensitive to changes in the electron density $n_e$ and temperature $T$, we have
shown by the gray shaded area, the case B predictions (normalized to unity at
Br 14) for  the parameter space of $n_e$ varying between $10^5$
to $10^{12}$cm$^{-3}$ and $T$ varying between 5000 to 20000K. By showing an extended range
in the parameter space, it becomes easier to establish when observed data clearly
falls outside the range of plausible model values. The second point
concerns the errors involved in comparing the luminosities of lines situated in
different
bands (Pa $\beta$ in $J$ band, Br 10-21 in $H$ band and Br $\gamma$ in
$K$ band). The robustness of the estimated line luminosities  depends
on how reliably the flux in the continuum has been calibrated in each of
these bands which in turn depends on how reliable is the estimate of the
reddening in the band concerned (the amount of extinction in each of the $J$,$H$ and
$K$ bands is different). Thus as a cross-check for the extinction of
$A_v$ = 2.26 obtained by Snijders  (1987), we note
that modelling of the galactic interstellar extinction by Marshall et al. (2006)
indicates a extinction in the direction of RS Oph of $A_k$ = 0.21 to 0.25
(or $A_v$ = 2.1 to 2.5 for $A_v$ = 10$A_k$) for an adopted distance of 1.6kpc
(Snijders 1987) to the
source. Thus there is a consistency in the $A_v$ values estimated by different
authors and it is unlikely that major errors are being introduced in the
analysis by an incorrect estimate of $A_v$. The third point concerns the
difficulty in estimating the equivalent widths (and hence line luminosities)
of the Br 10 and 11 lines because they are so strongly blended with the
Fe II lines. We have measured the equivalent widths of these lines after trying
to deblend the features using gaussian fits but the exercise is not entirely
satisfactory. Thus, while we do show the Br 10 and 11 lines  in Figure
6, their strengths are subject to considerable uncertainty (of about 30
percent) and they should be accepted with caution; the formal error in the
strength  of the other lines is in the range of 10 to 15 percent.

Examination of the spread in the Case B values in Figure 6 (the gray shaded area)
indicates that the higher Br lines (Br15 to 20) are really not sensitive enough to
discriminate whether there are deviations between the observed data
and Case B model values.  The lines that are sensitive  are Br $\gamma$ and Pa $\beta$.
In RS Oph, except for the 16 Feb 2006 data, it is seen that Br $\gamma$ is
consistently weaker than expected and that the
Br $\gamma$/Br14 ratio is generally in the range of 2 - 4 against an expected
median value of $\sim$ 6. There is also a suggestion that Br 10 and Br 11 are
stronger than expected and Br 10 nearly equal to or stronger than Br $\gamma$ on
all days except 16 Feb. Departures in the ratio of Pa $\beta$/Br $\gamma$ from
Case B are also seen - this ratio is expected to be  around 6 whereas we obtain
values of around   6.5, 9.5, 11.5,15,11 and 6 for the six days shown in Figure 6.
Thus the Pa $\beta$/Br $\gamma$ ratio diverges considerably from Case B on
most days except Feb 16 and  May 17. However, on May 17, the Br lines do
not conform to Case B behavior - the Br $\gamma$/Br14 ratio is low. The general
conclusion that is therefore suggested, after taking into account the behaviour
of all the HI lines, is that the Feb 16 data conforms to a reasonable extent
with Case B conditions; on other days there are marked deviations from it. On
these days, the observed behaviour of the Br lines
i.e the presence of a Br $\gamma$ line that is weaker than expected when compared to
the higher Br lines accompanied by the  additional observation
that the Br line strengths are possibly peaking around  Br 10 to 11 (instead of at
Br $\gamma$), suggests that the Br lines are optically thick. Such behaviour
of the Br lines  has been observed in
other novae like Nova Ophiuchi 1998  (Lynch et al. 2000) and V4643 Sgr (Ashok et
al. 2006) and also in Be stars (Banerjee et al. 2001). Such effects have been shown
by Lynch et al. (2000) to arise from  optical depth effects in an environment with
high particle densities ($n_e$ = $10^{10}$ to $10^{12}$cm$^{-3}$) which tends to
thermalize the level populations through collisions. It would thus appear that
invoking high density conditions could explain the observed deviations
from Case B. It is also noted that such high density conditions  prevailed throughout
our observations except in the very beginning at around 16 Feb 2006.  At this stage,
just four days after the outburst, the ejecta has been clearly shown to be in
a free expansion stage. That is, it had not yet swept up enough of the red giant
wind for the deceleration to begin and possibly for the dense contact surface of
the shockfront to fully develop. Hence, the high density conditions, that can
create a departure from Case B, were likely to be absent at this time.  The Fe II
line analysis of the previous sub-section, which also suggested the need for
high density conditions, is consistent with the line of argument  suggested here.

\subsection{Evolution of the coronal lines}

We consider the temporal evolution of the  [Si VI] 1.9641 $\mu$m and the [Mn XIV] 2.0894 $\mu$m coronal
lines here. The uncertainty in the assignment of the latter line to [Mn XIV]
has already been discussed in section 3.2; we proceed by assuming that this
assignment is correct. The observed strength of these lines on different days
is presented in Table 4 and their temporal evolution is shown in Figure 7.
The [Si VI] line is  first detected clearly on around Mar. 26.88 ($\sim$ 42 days after
outburst).
It then increases in strength, peaking at around March 30.86 ($\sim$ 46 days after outburst)
and subsequently
declines in intensity. The [Mn XIV] line appears on Mar. 2.95 ($\sim$ 18 days after outburst), earlier
than the [Si VI] line, but shows a similar evolution in the sense that it
increases in strength with time and subsequently declines. Unfortunately,
no spectra were recorded between 3 to 16 March 2006 (i.e. 18 to 32 days after
outburst) to monitor the evolution of the [Mn XIV] line during this period.
The observed temporal behaviour of the coronal lines can be interpreted in a
simple manner. The coronal lines are expected to arise from the hot, shocked
gas in the shockfront. This gas, whose temperature in the very early stages
was determined to be as high as $\sim$ $10^8$ K (Sokoloski et al. 2006), gradually
cools with time.  The fractional abundance of an ion (e.g. N(Si VI)/N(Si))
in a collisionally ionized, low density plasma is known to be temperature
dependent. Model calculations show that the fractional abundance of Si VI
peaks at  a temperature of 4$\times$$10^5$ K ( Jordan 1969; Shull
and van Steenberg, 1982) while the Mn XIV ion peaks at 1.6$\times$$10^6$ K
respectively (Landini and Monsignori Fossi, 1972). Since the
strength of a line associated with an ion is proportional to the abundance
of the ion,  it is expected that
the strength of the line will change as the temperature of the plasma changes
(thereby changing the abundance of the ion). It would appear that this
phenomenon is being manifested here. As the shocked gas in the RS Oph system
cools, a temperature will be reached that is conducive for the formation of
a  particular ion to begin. Subsequent lowering of the temperature should see
the fractional abundance of the ion  reach its peak value at the optimum
temperature for its production and  the fractional abundance should
subsequently decline as the temperature continues to decrease.

The temporal evolution of the [Si VI] and [Mn XIV] lines, as shown in Figure 7,  is
consistent with  such a scenario (there is, however,  a possible complication
which is discussed shortly) . Further, since the 2.0894 $\mu$m line is detected
earlier than the [Si VI] line (therefore at a phase when the shocked gas is
relatively hotter), the analysis  suggests that it is likely that the
2.0894 $\mu$m line originates from  an ion  with higher ionization potential than
Si VI.
In this sense it is consistent to identify this line with [Mn XIV] since it has
a higher ionization potential  of 344 eV compared to 166.8 eV for  [Si VI].  Since
the  [Mn XIV] and  [Si VI] line emission  peaks on March 16.95 and March 30.86
respectively,  the temperatures on these days are indicated to be 1.6$\times$$10^6$ K
and  4$\times$$10^5$ K   respectively based on the
ionization balance models discussed earlier ( Jordan 1969; Shull
and van Steenberg, 1982; Landini and Monsignori Fossi, 1972). These
temperatures may be compared with those derived from  X-ray observations on
nearby dates. Nelson et al. (2008) obtain a value of  6.3$\times$$10^6$ K on 11-12 March 2006 (day 27.7 after outburst; Table 4 of Nelson et al. 2008). This is reasonably in agreement with the temperature of  1.6$\times$$10^6$ K that we obtain on March 16.95.   Ness et al. (2007) determine a temperature range of  (6.3 - 8.3)$\times$$10^5$ K on day 39.7
after outburst (24 March 2006) while we get a value of 4$\times$$10^5$ K on March 30.86. Here too, the temperature estimates are  in reasonable agreement with each other.

\begin{figure}
\centering
\includegraphics[bb=1 0 415 542,width=3.0 in,height=4.2 in,clip]{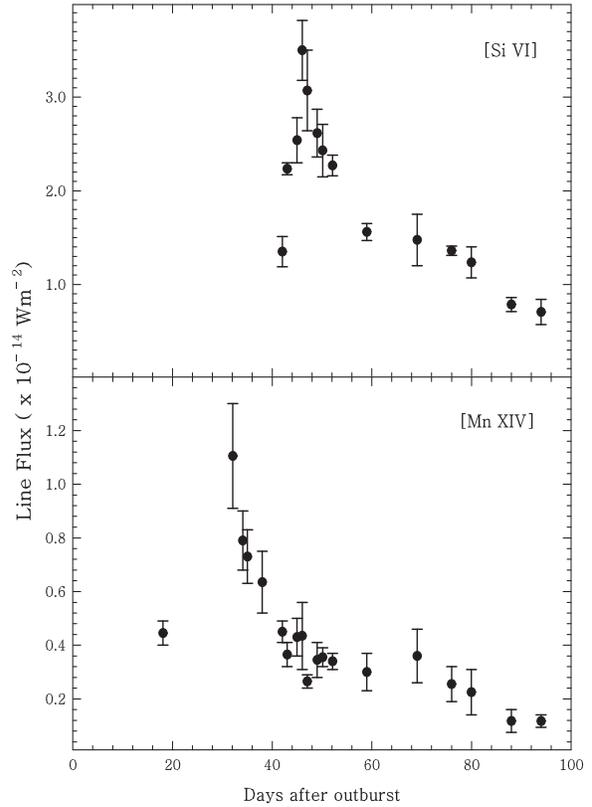}
\caption[]{ Temporal evolution of the de-reddened line strengths of the  [Si VI] 1.9641 $\mu$m and [Mn XIV] 2.0894 $\mu$m lines. Further details are given in section 3.7.}
\label{fig7}
\end{figure}

In the case of RS Oph, a complication that arises in applying  models for the
fractional ion abundance, based on purely collisional effects,
 is that the effect of any radiation field is assumed
 to be negligible. X-ray spectra in the initial stages after outburst  clearly indicate that the plasma in RS Oph is collisionally dominated (Nelson et al. 2008).
 However, these authors observe a super soft x-ray phase in RS Oph during weeks 6-10
 arising from the hot central white dwarf (WD)  whose temperature was
 estimated to be 800,000K.  The radiation field from the central WD could
thus be expected to affect the ionization balance in the ejecta. We consider it  a
difficult problem, beyond the scope of this work, to assess the extent
to which the WD radiation field - in conjunction  with collisional ionization - affects the ionization balance. We, however,
note that our analysis/discussion on the [Mn XIV] line is unlikely
to be affected by the super soft x-ray phase since this phase began  towards the end of March (Nelson et al. 2008) after the [Mn XIV] line had peaked in intensity.
However, the [Si VI] line emission, could be affected to some extent by the super-soft X-ray phase. 

\begin{table}
\caption[]{Evolution of the coronal lines}
\begin{tabular}{llcc}
\hline\\

Date of     &      Days after      & \multicolumn{2}{c} {Line flux ( x 10$^{-14}$ W m$^{-2}$)}   \\
Observation &      outburst      &      [Si VI]       &       	           [Mn XIV]		 \\

\hline
\hline \\

Mar. 2.9465	&	18.1165	     &	  	-	                &	0.445	$\pm$	0.045  \\
Mar. 16.9465	&	32.1165  &		-		            &	1.105	$\pm$	0.195 \\
Mar. 18.9372	&   34.1072	 &		-		            &	0.790	$\pm$   0.110 \\
Mar. 19.8833     &	35.0533	 &		-		            &	0.730	$\pm$	0.100 \\
Mar. 22.8682	&	38.0382  &		-		            &	0.635	$\pm$	0.115 \\
Mar. 26.8778     &	42.0478	 &	1.350	$\pm$	0.160	&	0.450	$\pm$	0.040 \\
Mar. 27.8644	&	43.0344	 &	2.235	$\pm$	0.065	&	0.365	$\pm$	0.045 \\
Mar. 29.8462	&	45.0162	 &	2.540	$\pm$	0.240	&	0.430	$\pm$	0.070 \\
Mar. 30.8629	&	46.0329  &	3.500	$\pm$	0.320	&	0.435	$\pm$	0.125 \\
Mar. 31.8701	&	47.0401	  &	3.070	$\pm$	0.430	&	0.265	$\pm$	0.025 \\
Apr. 2.8736	&	49.0436	  &	2.615	$\pm$	0.255	&	0.345	$\pm$	0.065 \\
Apr. 3.9691	&	50.1391	  &	2.430	$\pm$	0.280	&	0.355	$\pm$	0.035 \\
Apr. 5.9607	&	52.1307	  &	2.270	$\pm$	0.110	&	0.340	$\pm$	0.030 \\
Apr. 12.8454	&	59.0154	  &	1.560	$\pm$	0.090	&	0.300	$\pm$	0.070 \\
Apr. 22.9392	&	69.1092	  &	1.475	$\pm$	0.275	&	0.360	$\pm$	0.100 \\
Apr. 29.8355	&	76.0055	  &	1.360	$\pm$	0.050	&	0.255	$\pm$	0.065 \\
May 3.8100	    &	79.9800	  &	1.235	$\pm$	0.165	&	0.225	$\pm$	0.085 \\
May 11.8353	    &	88.0053	  &	0.785	$\pm$	0.075	&	0.118	$\pm$	0.043 \\
May 17.8103	    &	93.9803	  &	0.705	$\pm$	0.135	&	0.117	$\pm$	0.023 \\

\hline
\end{tabular}
\end{table}

\section{ Summary} $H$ and $K$ band photometry and spectroscopy of the 2006 outburst
of RS Ophiuchi are presented. This work, in conjunction  with the $J$ band
observations presented earlier in Das et al. (2006), documents the near-IR
evolution of RS Oph in a detailed manner.  An extensive set of spectra have
been presented and an analysis
and discussion has been made of the emission lines seen therein (HI Paschen and
Brackett recombination lines, coronal lines, Fe II lines etc.). The Fe II lines
at 1.6872 and 1.7414 $\mu$m, rather uncommon in  novae, have been studied in
some details. They appear to arise from a combination of Lyman alpha plus
Lyman continuum fluorescence and collisional excitation. The site of origin of
the near-IR line emission is explored based on the analysis of the Fe II lines
and Case B recombination studies. An analysis is made of the  temporal evolution of the [Si VI] 1.9641 $\mu$m coronal line and another coronal line at  2.0894 $\mu$m which is attributed to [Mn XIV].  Assuming collisional effects to dominate in the hot coronal gas, we estimate
the coronal temperature to be 1.6$\times$$10^6$ K
and  4$\times$$10^5$ K  on 2006 March 16.95 and March 30.86 respectively.

\section*{Acknowledgments}

The research work at Physical Research Laboratory  is funded by the Department
of Space, Government of India.  We are thankful for the online availability of
the Kurucz atomic linelist database which was extensively used in this work.
We thank the referee, Prof. Aneurin Evans, for his suggestions which
greatly helped in improving the results presented here.

\end{document}